\documentclass[aps,pre,superscriptaddress,reprint]{revtex4-1}
\usepackage{graphicx}
\usepackage[utf8]{inputenc}
\usepackage{amsmath}
\usepackage{amssymb}
\usepackage{hyperref}
\usepackage{url}
\usepackage{color}
\usepackage{cases}
\usepackage{xspace}

\newcommand\bigO[1]{\ensuremath{\OCAL (#1)}}

\newcommand\subcap[1]{{\textbf{#1}:}}

\newcommand{\SETOFM}{\MCAL}

\newcommand{\LJ}{\text{LJ}}
\newcommand{\Lfree}{L_\text{free}}
\newcommand{\HARM}{\text{harm}}
\newcommand{\FAC}{\text{fact}}
\newcommand{\FFeq}{\ensuremath{h^{\text{fact}}}}

\newcommand\subfig[2]{{Fig.~\ref{#1}{#2}}}

\usepackage{placeins}

\newcommand{\SET}[1]{\{#1\}}

\newcommand{\eq}[1]{eq.~\eqref{#1}}
\newcommand{\eqtwo}[2]{eqs~\eqref{#1} and~\eqref{#2}}

\newcommand{\Eq}[1]{Eq.~\eqref{#1}}
\newcommand{\fig}[1]{Fig.~\ref{#1}}
\newcommand{\figtwo}[2]{Figs~\ref{#1} and~\ref{#2}}
\newcommand{\quot}[1]{``#1''}

\newcommand{\sect}[1]{Section~\ref{#1}}

\newcommand{\MCAL}{\mathcal{M}}  %  mathcal
\newcommand{\OCAL}{\mathcal{O}}  %  mathcal
\newcommand{\expa}[1]{\mathrm{e}^{#1}}   % high exponential groupings a
\newcommand{\loga}[2][]{\log^{#1}\! \gla #2 \gra}  % log-brace,  with - nothing
\newcommand{\gla}{\,}  % ' group left a' 
\newcommand{\gra}{}  % ' group right a' 
\newcommand{\glb}{\left(}  % ' group left b' 
\newcommand{\grb}{\right)}  % ' group right b' 
\newcommand{\glc}{\left[}  % ' group left c'
\newcommand{\grc}{\right]}  % ' group right c' 
\newcommand{\TO}{,\ldots,}

\newcommand{\mean}[1]{\left\langle #1 \right\rangle}

\newcommand\bigOb[1]{\ensuremath{\OCAL\glb #1 \grb}}

\date{\today}
\begin{document}

\title{Event-chain Monte Carlo with factor fields}

\author{Ze Lei}
\email{ze.lei@ens.fr}
\affiliation{Laboratoire de Physique Statistique, Ecole Normale Sup\'{e}rieure
/ PSL Research University, UPMC, Universit\'{e} Paris Diderot, CNRS, 24 rue
Lhomond, 75005 Paris, France}

\author{Werner Krauth}
\email{werner.krauth@ens.fr}
\affiliation{Laboratoire de Physique Statistique, Ecole Normale Sup\'{e}rieure
/ PSL Research University, UPMC, Universit\'{e} Paris Diderot, CNRS, 24 rue
Lhomond, 75005 Paris, France}

\author{A.~C.~Maggs}
\email{anthony.maggs@espci.fr}
\affiliation{CNRS UMR7083, ESPCI Paris, PSL Research University, 10 rue
Vauquelin, 75005 Paris, France}

\begin{abstract}
We study the dynamics of one-dimensional (1D) interacting particles simulated
with the event-chain Monte Carlo algorithm (ECMC). We argue that previous
versions of the algorithm suffer from a mismatch in the factor potential between 
different particle pairs (factors) and show that in 1D models, this mismatch is
overcome by factor fields. ECMC with factor fields is motivated, in 1D,
for the harmonic model, and validated for the Lennard-Jones model as well
as for hard spheres. In 1D particle systems with short-range interactions,
autocorrelation times generally scale with the second power of the system size
for reversible Monte Carlo dynamics, and with its first power for regular
ECMC and for molecular-dynamics. We show, using numerical simulations,
that they grow only with the square root of the systems size for ECMC with
factor fields. Mixing times, which bound the time to reach equilibrium from an
arbitrary initial configuration, grow with the first power of the
system size. 
\end{abstract}
\maketitle 

%\tableofcontents
\section{Introduction}
The dynamics of physical systems plays an important role in numerous fields of
science.  The study of dynamics aims at elucidating equilibrium and
non-equilibrium phenomena, including correlation functions, 
coarsening dynamics after a quench, and manifestly
non-equilibrium phenomena such as turbulence.  In computational statistical
physics, Markov chain Monte Carlo (MCMC)~\cite{Metropolis1953,Levin2008,SMAC}
and molecular-dynamics (MD) algorithms~\cite{Alder1957} are often employed to
generate equilibrated samples and to determine thermodynamic averages and
correlations.  The non-equilibrium aspect then consists in characterizing the
approach to equilibrium from an arbitrary, atypical, initial condition.  This is
quantified by the mixing time~\cite{Levin2008}, an important figure of merit for
MCMC. The other important time scale characteristic of a physical system 
is the autocorrelation times of the underlying Markov process given by the 
inverse gap of the transition matrix. 

In  reversible Markov chains (as used in the vast majority of MCMC algorithms),
the requirement that the long-time steady state corresponds to thermodynamic
equilibrium is expressed through the detailed-balance condition, which assures
that all the net probability flows vanish in equilibrium. In recent years,
however, irreversible Markov chains have been found to show considerable
promise 
\cite{Turitsyn2011,SakaiIsing2013,SakaiEigenvaluePRE2016,Bierkens2018}. They 
feature a steady state with non-vanishing net probability flows
if a weaker global-balance condition is satisfied.  Global balance corresponds
to an incompressibility condition in configuration space: the steady-state
flows into each configuration sum to the flows out of it.

An example of an irreversible Markov chain is the event-chain algorithm
(ECMC)~\cite{Bernard2009,Michel2014JCP}. This algorithm has been
successfully applied to many problems from hard-sphere and soft-sphere
melting~\cite{Bernard2011,Engel2013,Kapfer2015PRL} to spin 
models~\cite{MichelMayerKrauth2015,Nishikawa2015} and quantum-field 
theory~\cite{HasenbuschSchaefer2018}. In this paper, we study the relaxation 
times of ECMC in one-dimensional (1D) models of $N$ particles with local 
interactions~\cite{KapferKrauth2017,Lei2018b}.  We analyze in detail the 
% \FIXME{Ze: Please update the reference... it's still a preprint}
relaxation of both Lennard-Jones and hard-sphere models, study the statistical 
properties of ECMC trajectories and show how to greatly accelerate known 
algorithms by the introduction of a \quot{factor field}, which compensates the 
system pressure, $P$, without influencing physical properties. 

\subsection{Characteristic times of  Markov chains}

Irreversible MCMC algorithms can be faster than their reversible counterparts.
A particularly interesting case is the 1D hard-sphere model of $N$ spheres
 (rods).  For this model, the local heat-bath algorithm mixes in 
\bigOb{N^3\log{N}} moves~\cite{RandallWinklerInterval2005} on an interval with 
fixed boundary
conditions. The mixing time for the same model with periodic boundary conditions
is between \bigOb{N^3} and \bigOb{N^3\log{N}}~\cite{RandallWinklerCircle2005}:
Simulations favor the latter~\cite{KapferKrauth2017}. The reversible Metropolis
algorithm has a similar mixing behavior. Various local irreversible Markov
 chains mix in \bigOb{N^{5/2}} moves (forward Metropolis 
algorithm~\cite{KapferKrauth2017}), in \bigOb{N^{2} \log{N}} steps (lifted 
forward
Metropolis algorithm and ECMC~\cite{KapferKrauth2017,Lei2018b}) and even
\bigOb{N^{2}} single moves with a re-labeling ECMC~\cite{Lei2018b}.

The lifted forward Metropolis algorithm in continuous space with infinitesimal
movements constitutes ECMC. For hard spheres, it is deterministic without 
restarts, but then mixes in $\bigOb{N^2\log N}$ steps at randomized stopping 
times~\cite{Lei2018b}.  Although ECMC is irreversible under a transformation
of the time $t \mapsto -t$, under the combined transformation of times and 
positions $(t, x) \mapsto (-t, -x)$, the dynamics runs backwards in time. 
The irreducibility of the  lifted forward Metropolis algorithm can be shown 
using this time-reversal property. It may also explain  why ECMC is typically as 
fast as MD.

Previous work has also explored the autocorrelation times (rather than the
mixing times) under ECMC dynamics in the $D$-dimensional harmonic-solid model,
of which the equilibrium properties can also be obtained exactly 
(see~\cite{Lei2018}). In 1D, the dynamic exponent of the autocorrelations under 
ECMC
dynamics takes the strikingly low value of $z=1/2$, corresponding to an
equilibrium autocorrelation time involving \bigO{N^{3/2}}  moves or
$\tau \sim \bigO{N^{1/2}}$ {sweeps}. This is $N^{1/2}$ times smaller
(faster) than the best autocorrelation time found in the hard-sphere
system.

The present paper starts from the similarity between the dynamics of the 1D
harmonic-solid model and that of the Lennard-Jones model at low temperature
$T$. We generalize this favorable scaling of the harmonic model  to all 
$T$ in the Lennard-Jones model as well as the hard-sphere model. We
expect that this concept can also be generalized for higher-dimensional 
models~\cite{Faulkner2018,MaggsMultiscale2006}.

\subsection{1D particle systems, algorithms}\label{sect:syst-algo}

We consider a 1D system of $N$ particles $i\in \SET{1 \TO N}$ with $x_i < 
x_{i+1}$ on an interval of 
length $L$ with periodic boundary conditions in $N$ and in $L$. 
In the reversible local Metropolis algorithm, at each iteration, a
randomly chosen particle $i$ is proposed to move as $x_i \to
x_i + \text{ran}[-\epsilon, \epsilon]$, where $\text{ran}$ is a random number
uniformly distributed between $-\epsilon$ and $\epsilon$.  For hard spheres,
the move is accepted if the new sphere position does not lead to overlaps
with spheres $i-1$ and $i+1$, and in addition does not induce a change of the
ordering.  In the presence of a potential $U$, the
move is accepted with probability  $\min(1, \exp(- \Delta U / T))$, where
$\Delta U$ is the change in potential for the proposed move.  The amplitude
$\epsilon$ is chosen to maximize the speed of the method. In the heat-bath
algorithm the distribution of the particle $i$ is fully resampled in the
potential of its neighbors at each time step.

ECMC, for one-dimensional hard spheres \cite{Bernard2009,KapferKrauth2017}, 
consists in moving spheres in a chain-like manner. Up to a restart time, 
sphere $i$ moves with unit velocity until it collides with sphere $i+1$, at 
which moment it stops, and sphere $i+1$ moves forward. For each of the 
subsequent \quot{chains} (the displacements between restarts), the starting 
sphere is randomly chosen, and the length of the chain (the time until the next 
restart) is sampled from a distribution on the length scale $L$. 
For a more general interaction potential, ECMC breaks up the 
total system potential up into separate \quot{factor potentials}, each 
of which is treated independently \cite{Michel2014JCP,Peters_2012}. A factor 
potential provides for a randomized stopping time. For a given move
involving particle $i$, the 
smallest stopping time of all factors provides the next event time. 
The next particle to move is determined through a lifting scheme 
\cite{Harland2017,Faulkner2018} from the factor
triggering the event. With potentials more general than hard spheres, restarts 
are no longer required to ensure irreducibility of ECMC.

In ECMC, path statistics in equilibrium and pressure $P$ are linked by 
\begin{equation}
    P / T \propto  
\frac{1}{t}    \mean{x_{i(t)} - x_{i(0)} },
\label{equ:ecmcPressure}
\end{equation}
where $x_i(t)$ is the position of the particle that is active at time
$t$~\cite[eq.(20)]{Michel2014JCP}.
\Eq{equ:ecmcPressure} holds for all time intervals $t$.  It 
is very convenient as an unbiased estimator of the pressure, and has been much 
used~\cite{Kapfer2015PRL}. The factor fields of the present paper will allow us 
to exactly compensate the pressure without affecting the physical properties of 
the system, and lead to greatly accelerated ECMC methods.

In many-particle simulations, MD algorithms generally feature smaller
relaxation time scales than MCMC methods. In essence this is because
momentum conservation (present in MD, but absent in MCMC)  allows for
faster transport of inhomogeneities in the velocity and position fields
(\cite{AlderWainwrightAutocorrelation1970,WittmerPolinska2011}).
In our comparisons with ECMC, MD simulations are performed
using the leapfrog or Verlet algorithm coupled to a Langevin thermostat for
the velocity. The integration time step $\delta t$ is adjusted by finding the 
stability limit of the integrator, then reducing $\delta t$ by 
an order of magnitude.
Inverse error analysis shows that the effective Hamiltonian is close to
that of the original model, with a systematic shift of $O(\delta t^2)$
in the effective Hamiltonian~\cite{Hairer2006}.  We choose the strength of 
coupling to the Langevin
thermostat so that the longest wavelength mode is close to critically damped.

We concentrate our measurements on the dynamics of the structure factor of the
lowest Fourier coefficient
\begin{equation}
 S({q}) = \frac{1}{N}\left|\sum_{j=1}^N\expa{i {q}\cdot x_j} 
\right|^2,
 \label{equ:structure-factor}
\end{equation}
with $q=2 \pi/L$, which is sensitive to large-scale motion of particles. The
integrated autocorrelation times $\tau$ of $S(\frac{2\pi}{L})$ are measured in
\quot{sweeps}, that is, a constant time interval for all $N$ particles in MD,
$N$ attempted displacements in MCMC, or $N$ events in ECMC (in comparing the
methods, we compensate for the different implementation speeds of a sweep in MD,
MCMC, ECMC). We use the blocking method~\cite{error} to quantify the algorithm
speed.

\subsection{ECMC for harmonic interactions}\label{sect:motiv-harm}

We first consider a harmonic potential with a minimum at a separation
$b$
between neighboring particles:
\begin{equation}
 U_\HARM(x_{i+1} - x_i; b) = \frac{k}{2}\glc (x_{i+1} - x_i) - b \grc ^2,
\end{equation}
where periodic boundary conditions for the particle separation $x_{i+1} -x_i$
are taken. They are also implied for the particle indices. 
The total potential of the system of a fixed length $L$ is
\begin{align}
  U_\HARM(\SET{x_i};b) =& \frac{k}{2} \sum_{i=1}^{N} {(x_{i+1} - x_i - b)}^2 
\label{equ:first-harmonic} \\
=& U_\HARM(\SET{x_i};0) - kbL + \frac{1}{2}Nkb^2,
\label{equ:shift-harmonic}
\end{align} 
where periodic boundary conditions in $N$ and $L$ are again understood.
Because of the periodic boundary conditions, 
the choice of the equilibrium separation $b$ simply
shifts the ground-state potential, without changing the stationary distribution
and equilibrium correlations. Nevertheless, the ground-state potential is
dependent on $L$ and it determines some thermodynamic properties, such as the
pressure:
\begin{equation}
 P_{\HARM}(b) = k (b - L/N).
 \label{equ:P-harm}
\end{equation}
The system with $b=L/N$ satisfies $P_\HARM=0$.

\begin{figure}[htbp]
\includegraphics[width=0.8\linewidth]{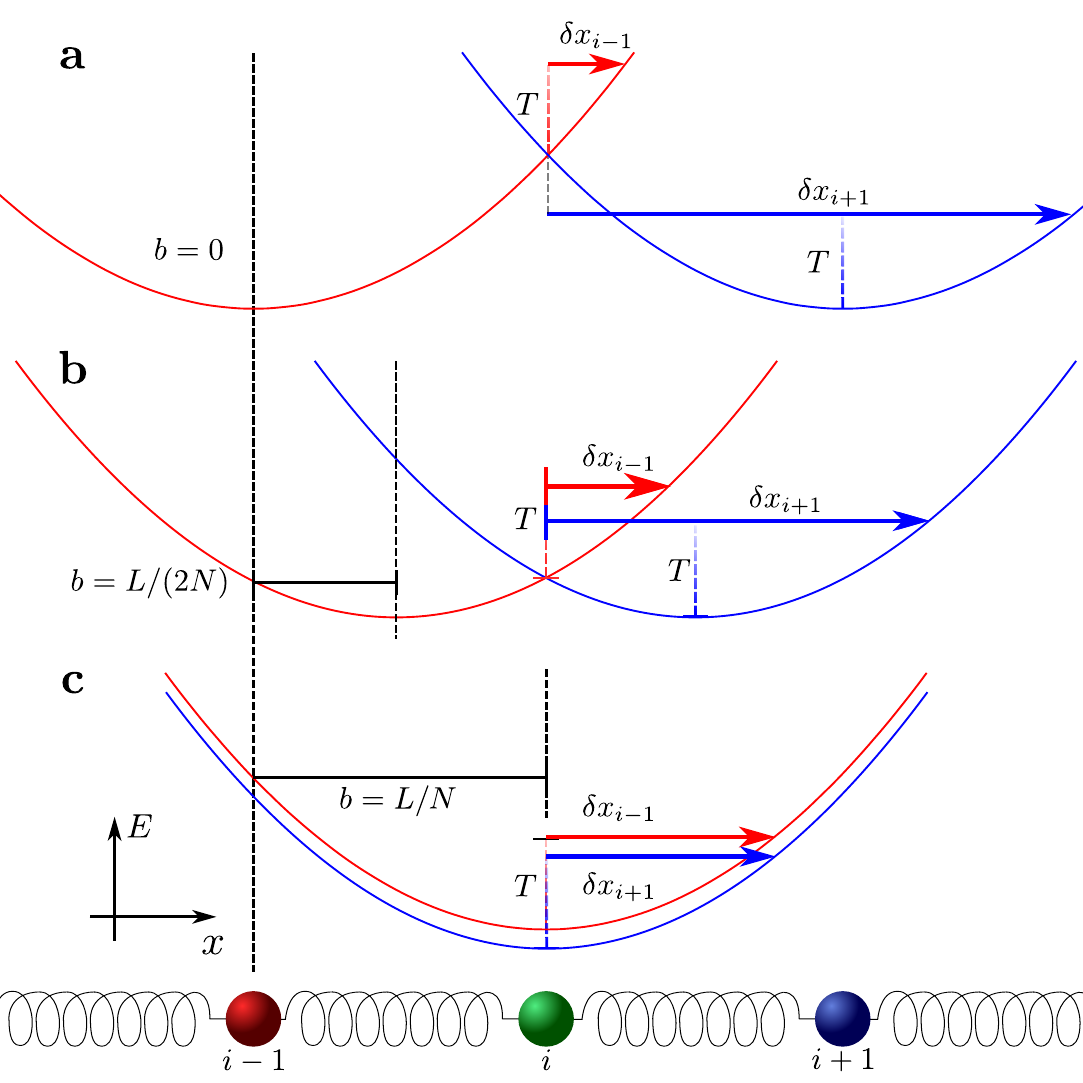}
\caption{ECMC dynamics in the harmonic model. Particle $i$ is active,
and the red and blue curves indicate the interactions with 
particles $i-1$ and $i+1$, respectively.
The proposed moves are indicated by solid colored
arrows, with respect to the temperature $T$ (vertical dashed lines).  \subcap{a}
For $b=L/N$, the interractions with particles $i-1$ and $i+1$ overlap. 
\subcap{b} For $b=L/(2N)$, the
potential with $i-1$ proposes a smaller displacement than $i+1$.
Particle $i-1$ is more likely to
be activated.  \subcap{c} For $b=0$, the moves proposed by  $i-1$
and by $i+1$ are balanced.
}\label{fig:Harmonic}
\end{figure}

In a periodic system, 
MD and the reversible Metropolis algorithm are strictly 
independent of $b$, as they only rely on the forces (identical derivatives of 
\eqtwo{equ:first-harmonic}{equ:shift-harmonic} with respect to the $x_i$) or  
potential differences between configurations. 
However, the explicit form of the pairwise interaction influences the 
ECMC dynamics, as the factor potentials are treated independently. One such 
factor potential may thus contain the single term $U_\HARM(x_{i+1} - x_i; b)$ 
with its explicit dependence on $b$.
In the following we consider such factors between all 
neighboring pairs of particles. 
For $b=0$, the harmonic interactions on particle $i$ from its
neighbors is attractive if $x_{i-1} < x_i < x_{i+1}$. It implies that for an
active particle $i$ with a positive displacement, the particle $i-1$ is 
likely to trigger the next event in ECMC (and to be the next active particle) 
(see \subfig{fig:Harmonic}{a}). The displacement $\delta x$ per event is:
\begin{equation}
\delta x \sim \frac{T}{k (L/N-b)}, \;\text{if }T\ll \frac k 2 {(L/N-b)}^2.
\label{equ:harmonic_b}
\end{equation}
As $b$ increases, the triggering probability is less biased and the
displacement gets larger, and eventually reaches the maximum:
\begin{equation}
\delta x \sim \sqrt{\frac{2T}{k}},  \;\text{when }b=L/N,
\label{equ:harmonic_0}
\end{equation}
with  symmetric triggering probabilities in both directions
(see \subfig{fig:Harmonic}{c}).

At low $T$, the displacement per event in \eq{equ:harmonic_b} is much smaller
than that in \eq{equ:harmonic_0}. We expect that the case $b=L/N$ leads to 
larger amplitude movements of the active particle $i$, at the same time the 
transfer of activity is equally often toward $i+1$ and toward $i-1$, and 
characterizes the detailed ECMC dynamics. The case $b=L/N$ indeed gives rise 
to the exceptionally fast dynamics, characterized by $z=1/2$~\cite{Lei2018}. 
The aim of the present paper is to generalize this fast relaxation to arbitrary 
potentials. 

\section{Linear Lennard-Jones model with ECMC} 

We study now the Lennard-Jones potential
\begin{equation}
  U_\text{LJ}(r) = \frac{1}{r^{12}} - \frac{1}{r^6},
  \label{equ:LJ-potential}
\end{equation}
where $r= x_{i+1} - x_i$, with periodic booundary conditions. The minimum
is $U_{\LJ}(r_{\min}) = -1/4$ for  $r_{\min}=2^{1/6}$, which sets a typical 
potential
scale of the system $\epsilon = |U_{\LJ}(r_{\min})|$.  In previous 
work~\cite{Faulkner2018} we proposed multiple factor sets within ECMC. Here we 
take into consideration two distinct factor sets $\SETOFM_{LJ}$ and 
$\SETOFM_{6+12}$:
the former groups the terms $1/r^6$ and $1/r^{12}$ into a single Lennard-Jones
factor, while the latter treats them separately as two factors which
independently trigger events.

As the particles always move in the positive direction ($x_i$ is always
increasing), the active particle $i$, with the factor set $\SETOFM_{6+12}$, will
either trigger the particle $(i+1)$ by the repulsive contribution $1/r^{12}$
or the particle $(i-1)$ by the attractive contribution $1/r^{6}$. The factor
set $\SETOFM_{LJ}$ can lead to a trigger from $(i+1)$ or $(i-1)$ since the
Lennard-Jones interaction has both increasing and decreasing branches.

In the following, we will show that the large-scale dynamics of ECMC are
very sensitive to the choice of factor sets, even if all choices lead to
the same equilibrium state. Good choices are crucial in the creation of
efficient algorithms.

\subsection{Simulations of 1D Lennard-Jones models}

\begin{figure}[htbp]
  \includegraphics[width=\linewidth]{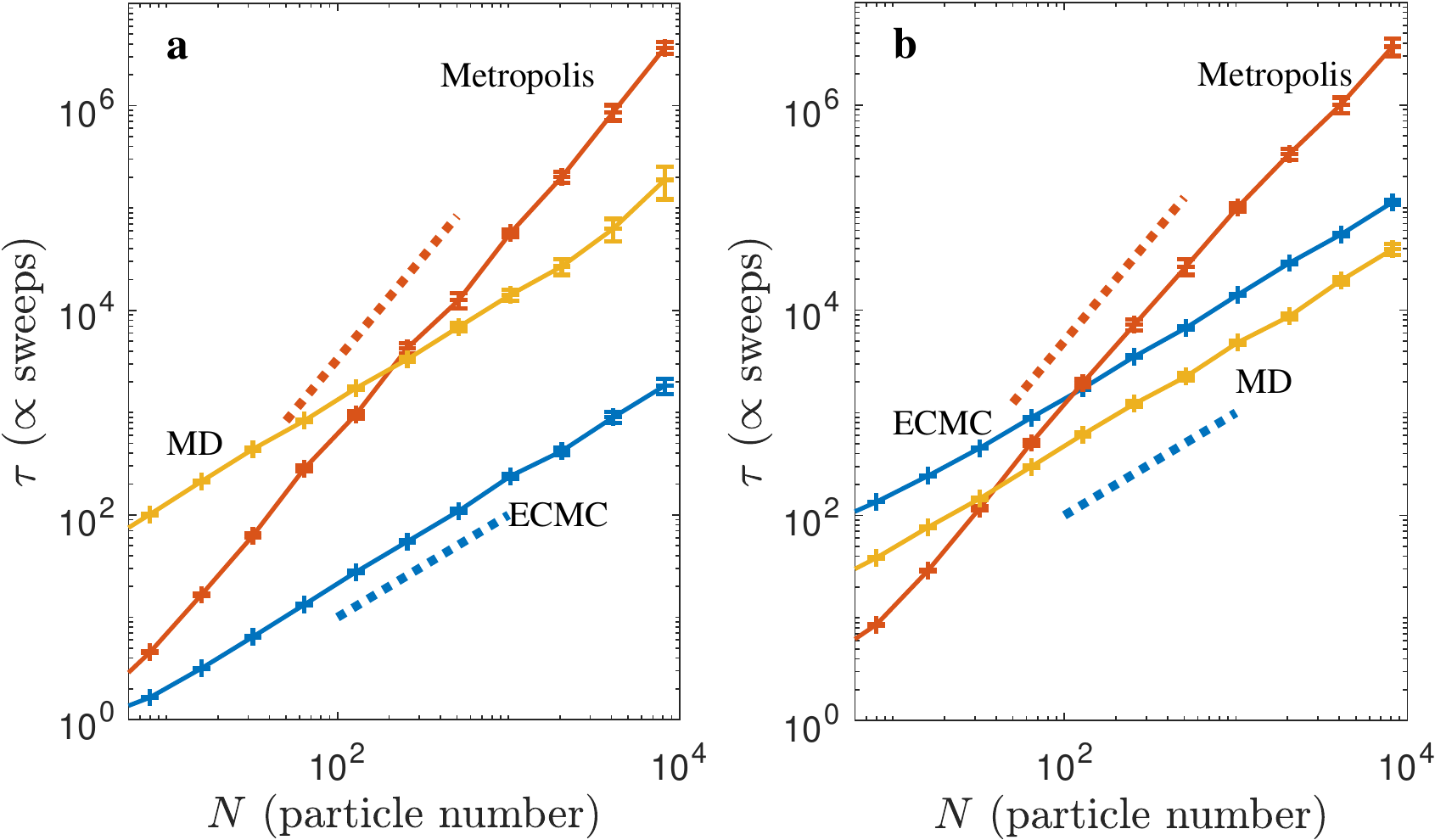}
  \caption{Equilibrium autocorrelation times $\tau$ vs.\ system size $N$, for 
   the 1D
   periodic Lennard-Jones model.  Reversible local Markov-chain dynamics, ECMC
   with restarts and MD. \subcap{a} $T / \epsilon=10$, combined factors,
   $\SETOFM_{LJ}$. \subcap{b} $T /  \epsilon=0.1$, separate factors,
   $\SETOFM_{6+12}$. Scalings $\tau \sim N$ and $\tau \sim N^2$ are indicated
   with dotted lines.}\label{fig:P1L}
\end{figure}

We simulate a slightly compressed ($P>0$) linear Lennard-Jones
model with periodic boundary conditions with average separation between particles
equal to $\Delta=1.06<r_{\min}$ and use the reversible local Metropolis 
MCMC method, MD, as well as ECMC with the factor set $\SETOFM_{LJ}$
(see \subfig{fig:P1L}{a}). Metropolis MCMC is asymptotically the slowest 
method
for $N \to \infty$: the autocorrelation time (measured in sweeps) increases as
$N^z$ with $z=2$ characteristic of the diffusion of density
fluctuations. MD is better behaved, due to the propagative
compressional waves which more efficiently sample long-wavelength
modes. MD is however disadvantaged by the necessity of using a
small integration time step $\delta t$ to stably explore the dynamics. The 
result from ECMC
is very favorable, we see a low dynamic exponent ($z=1$) combined with a small
prefactor in the scaling: the algorithm makes a large leap (without systematic
errors) for each iteration.

However, ECMC can also be less efficient than MD, in certain implementations 
(see \subfig{fig:P1L}{b}). Here we use the factor set
$\SETOFM_{6+12}$, at low $T$. Here an analogous phenomenon occurs to that
displayed in \fig{fig:Harmonic}, in a form which is amplified by the splitting
of the $1/r^6$ and $1/r^{12}$ contributions to the potential. The algorithm
advances with the use of steps which are too small to efficiently explore the
local environment.  This slowdown of ECMC at low $T$ was pointed 
out previously~\cite{Hu2018}. We now study analytically the Lennard-Jones 
interaction in 
\eq{equ:LJ-potential} at low $T$, and make contact with the harmonic model in 
order to eliminate this slowdown. 

\begin{figure}
  \includegraphics[width=0.99\linewidth]{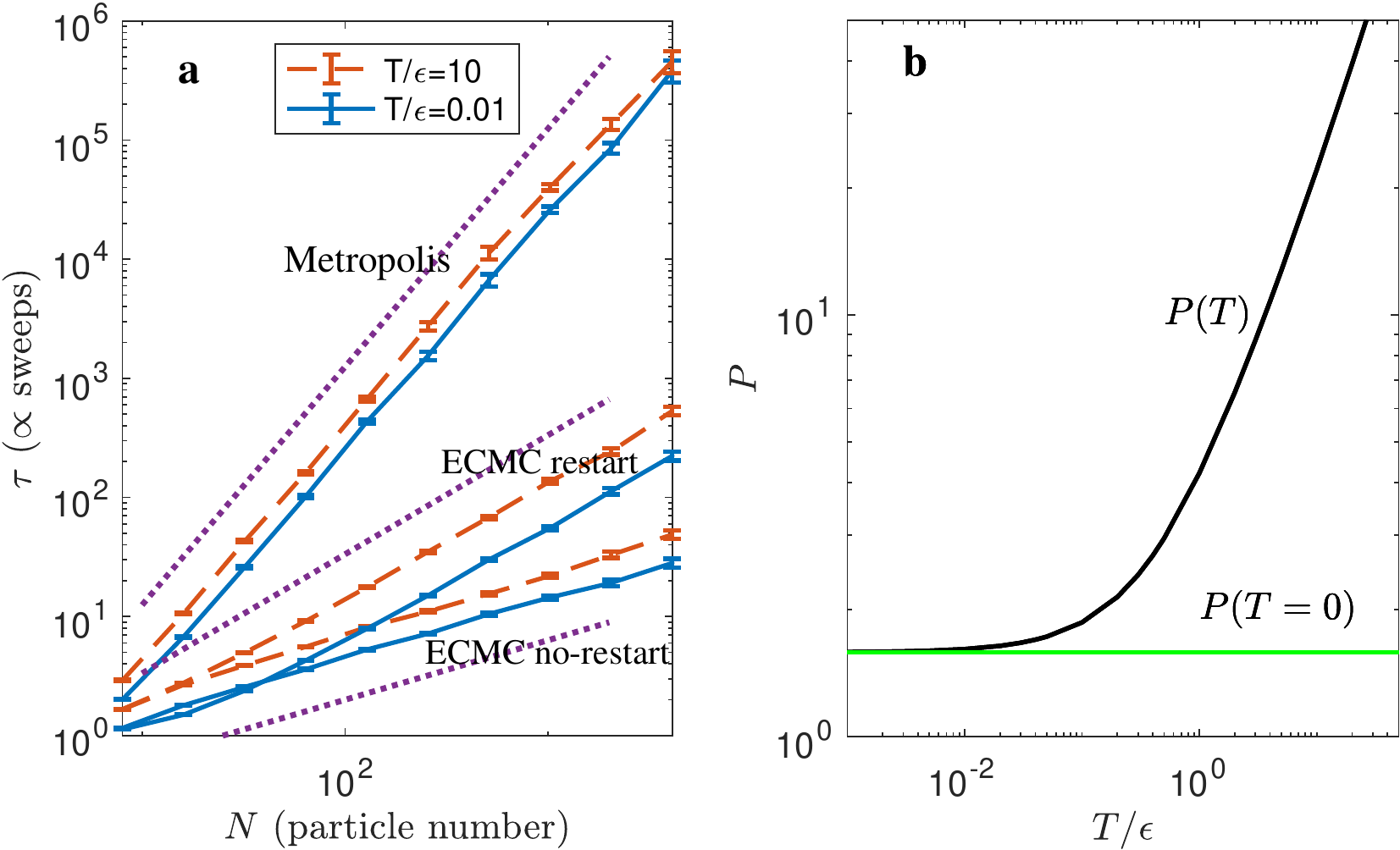}
  \caption{Equilibrium autocorrelation timess $\tau$ vs.\ system size $N$,  for 
   the 1D periodic Lennard-Jones system
   with factor fields.  \subcap{a} Reversible local Metropolis  dynamics and
   ECMC with and without restarts, at
   high $T$, $T / \epsilon=10$ and low $T$, $T / \epsilon=0.01$.  
Scalings
   $\tau \sim N^{1/2}$, $\tau \sim N$, and $\tau \sim N^2$ are indicated with
   dotted lines.  \subcap{b} Pressure $P(T)$ and its $T\to 0$ limit (see
   \eq{equ:hmoleq1}).}\label{fig:LJ_CMF}
\end{figure}

A straightforward expansion of the potential $U_{\LJ}(r)$ of 
\eq{equ:LJ-potential} to second order around a generic position $r=\Delta$ 
yields
\begin{align}
  U_{\LJ}(r) &=  U_{\LJ}(\Delta) - h_\text{LJ}(\Delta) (r - \Delta) \\ 
	     &+ \frac{1}{2} k_\text{LJ}(\Delta) {(r - \Delta)} ^ 2 + \ldots. 
\label{equ:LennardJonesExpansion}  
\end{align} 
This validates the (obvious) fact that the 1D Lennard-Jones model, in the limit
$T \to 0$, is described by a harmonic model % with ($L/N<2^{1/6}$) 
with, in analogy to \eq{equ:shift-harmonic}, a \quot{stiffness}
\begin{equation}
  \frac{1}{2}k_{\text{LJ}}(\Delta) =  \glb  \frac{78}{ \Delta^{14}} - 
  \frac{21}{\Delta^8}  \grb
\end{equation}
and a linear coefficient
\begin{equation}
  h_\LJ(\Delta) =  - 6(-2 + \Delta^6)/\Delta^{13}.
  \label{equ:hmoleq1}
\end{equation}
Summed over the $N$ pairs $(i,i+1)$ (with periodic boundary conditions), the 
constant ($U_\LJ(\Delta)$) and the first-order term in 
\eq{equ:LennardJonesExpansion} are without incidence on the 
constant-volume thermodynamics and the stationary distribution.

Analogously to \eq{equ:shift-harmonic}, we may add a 
temperature-dependent factor-field interaction
\begin{equation}
U^{\FAC}_\LJ = h^\FAC_\LJ(T) \sum_{i=1}^N  \glb x_{i+1} - x_{i} \grb
\end{equation}
to the total Lennard-Jones potential $\sum_{i}U_\LJ(x_{i+1} - x_i)$.
The model defined by $U_\LJ + U^{\FAC}_\LJ$ differs from the model given by 
$U_\LJ$ alone (in the presence of periodic boundary conditions), as the two 
have different pressures. Nevertheless, the samples obtained from the two 
associated Boltzmann distributions are the same, 
and therefore also all probability distributions and correlation 
functions at constant $L$. 
We choose a factor field to exactly compensate the linear term in the
interaction in the limit $T \to 0$:
\begin{equation}
  \FFeq_\LJ(T) = h_\LJ(\Delta) \quad \text{(for small $T$)}.
  \label{equ:hmoleq2}
\end{equation}
This clearly eliminates the inefficiencies of ECMC at low temperatures.
In the model defined by \eq{equ:hmoleq2}, the pressure $P$ vanishes as $T\to 
0$. 
Because of the connection between the pressure and the path statistics 
expressed in \eq{equ:ecmcPressure}, the ECMC trajectories are then without a 
drift term, and the expected displacement vanishes. 
As we now confirm numerically, we can speed up ECMC at arbitrary $T$ by 
adopting a factor potential that exactly compensates for $P$.

We start by performing a set of short simulations to measure $P$
from \eq{equ:ecmcPressure} (see \subfig{fig:LJ_CMF}{b}). The function 
$P(T)$, thus obtained, recovers the 
$T\to 0$ limit. We then fix the value of the factor field
in longer simulations to characterize the dynamics (see \subfig{fig:LJ_CMF}{a}). 
Indeed, both at low and at high $T$, ECMC remains efficient, 
and the dynamical exponent $z = 1/2$ corresponds to the
harmonic model for $b = L/N$. This was tested for temperatures as high as
$T / \epsilon=10$ where the interactions for Lennard-Jones particles are
dominated by the short-ranged repulsive core. The ansatz $\FFeq=P$ for the 
factor field thus holds at temperatures at which the harmonic approximation 
of the potential no longer applies. Maximum efficiency is found for ECMC without 
restarts only: restarting the chain after $\sim N$ events leads to a larger 
dynamic 
exponent. 

For the Lennard-Jones system, ECMC with factor fields requires 
finding roots to the equations
\begin{equation}
  \frac{1}{r^{12}} - \frac{1}{r^6}  \pm P r = \Delta U.
  \label{equ:iterate}
\end{equation}
We use the iterative Halley method~\cite{Halley}, a higher-order 
generalization of  Newton's method. It has the advantage of 
stability when starting an iteration near a stationary point of the function
\eq{equ:iterate}. We start the iteration with a guess obtained with one of two
methods. For small $\Delta E$ we make a harmonic approximation to the left-hand
side of \eq{equ:iterate}.  For large $\Delta E$, the starting point
is approximated
as a root to the equation $1/r^{12} = \Delta E$. The iteration converges to
machine precision within three iterations. The relative speeds shown in
\subfig{fig:LJ_CMF}{a} account for this slow, iterative step through an 
appropriate proportionality factor for each algorithm.
Alternatives to root finding may including thinning methods 
(as used in  the cell-veto
algorithm~\cite{KapferKrauth2016}) which compare rates derived from
\eq{equ:iterate} to an analytically tractable bound.

\subsection{Extensions: Alternative factor sets}

ECMC allows for many other choices for factors and also for
lifting schemes. We may generalize the factor field method to 
the $\SETOFM_{6+12}$
factor set by introducing one  factor field each for $1/r^6$ and for $1/r^{12}$
interactions (checking the convergence of the method for multiple correlation
functions), but we did not explore fully the optimal choice of the two 
factor fields. We also studied factor sets which contain all the interactions of 
the model. This scheme is particularly interesting because the active particle 
$i$ simultaneously explores the potential due to both $i-1$ and $i+1$, 
without the need for an explicit factor field. Again, 
this scheme uses an iterative solver. The full system
factors also require a more sophisticated lifting scheme --
generalizations of the \quot{inside first} and \quot{outside first} methods
~\cite{Faulkner2018}. Particles with positive factor derivatives
and particles with negative factor derivative are aligned in index order (see
for instance~\cite[Fig.~10]{Faulkner2018}). The lifting dynamics corresponds
to the
alignment of factors vertically. In such schemes, factors contain $\bigOb{N}$
terms.  Efficient alignment of the lifting diagram requires the use of a tree
structure for bookkeeping with an effort $\bigOb{\loga{(N)}}$. We found this 
method however to be less efficient than the factor field, and so do not report 
further on speed measurements.

\section{Factor fields for 1D hard spheres}
\begin{figure}[htbp]
\includegraphics[width=0.7\linewidth]{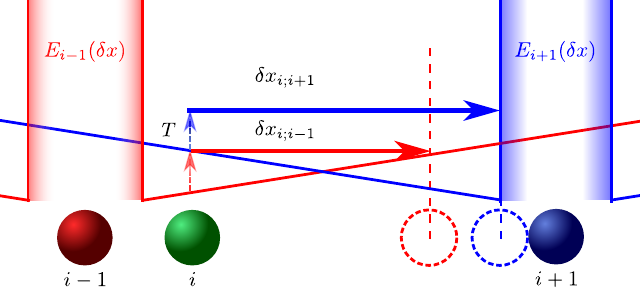}
\caption{ECMC dynamics with factor fields (indicated by inclined straight 
lines) 
for 1D 
hard spheres. The moves of the active sphere $i$ 
proposed by the factors with $i+1$ and $i-1$
are indicated by horizontal arrows and dashed sphere positions. In the 
optimal dynamics, the slopes of the factor fields equal $\pm P$ in 
\eq{equ:tonks}.
}\label{fig:HardspheresSchema}
\end{figure}

To illustrate the generality of factor fields, we now consider their
application to the 1D hard-sphere model, where the potential can no longer be
expanded as a power series (as in \eq{equ:LennardJonesExpansion}). 
Nevertheless, 
the model has a well-defined pressure, that is computed from the partition
function $ Z = {(L-N\sigma)} ^N$. This gives for the free energy, at temperature
$T$, $ F = -T \log Z = -N T \log \Lfree$, with $\Lfree = L - N\sigma$ and 
therefore~\cite{Tonks1936}
\begin{equation}
P = \frac{NT}{\Lfree}.
\label{equ:tonks}
\end{equation}

\subsection{Implementation, autocorrelation times}
The implementation of ECMC with factor fields for hard spheres does not require
numerical root finding: an active particle $i$, moving to the right,
generates two possible events, a hard-sphere collision with the particle
$i+1$ or else a trigger due to the  factor field of particle $i-1$ (see
\fig{fig:HardspheresSchema}).  The latter path length is sampled from an
exponential distribution
\begin{equation}
  \rho(x) = \frac{P }{T} \expa{-x P/T }.
\end{equation}
The smaller of the two proposed displacements yields  the next event, and it 
defines the lifting, as the new active particle is the 
one that has triggered the event.  Irreducibility is
guaranteed in the dynamics with an infinite event chain, and restarts are no
longer needed, unlike for hard-sphere ECMC without the factor field.

\begin{figure}[htbp]
\includegraphics[width=\linewidth]{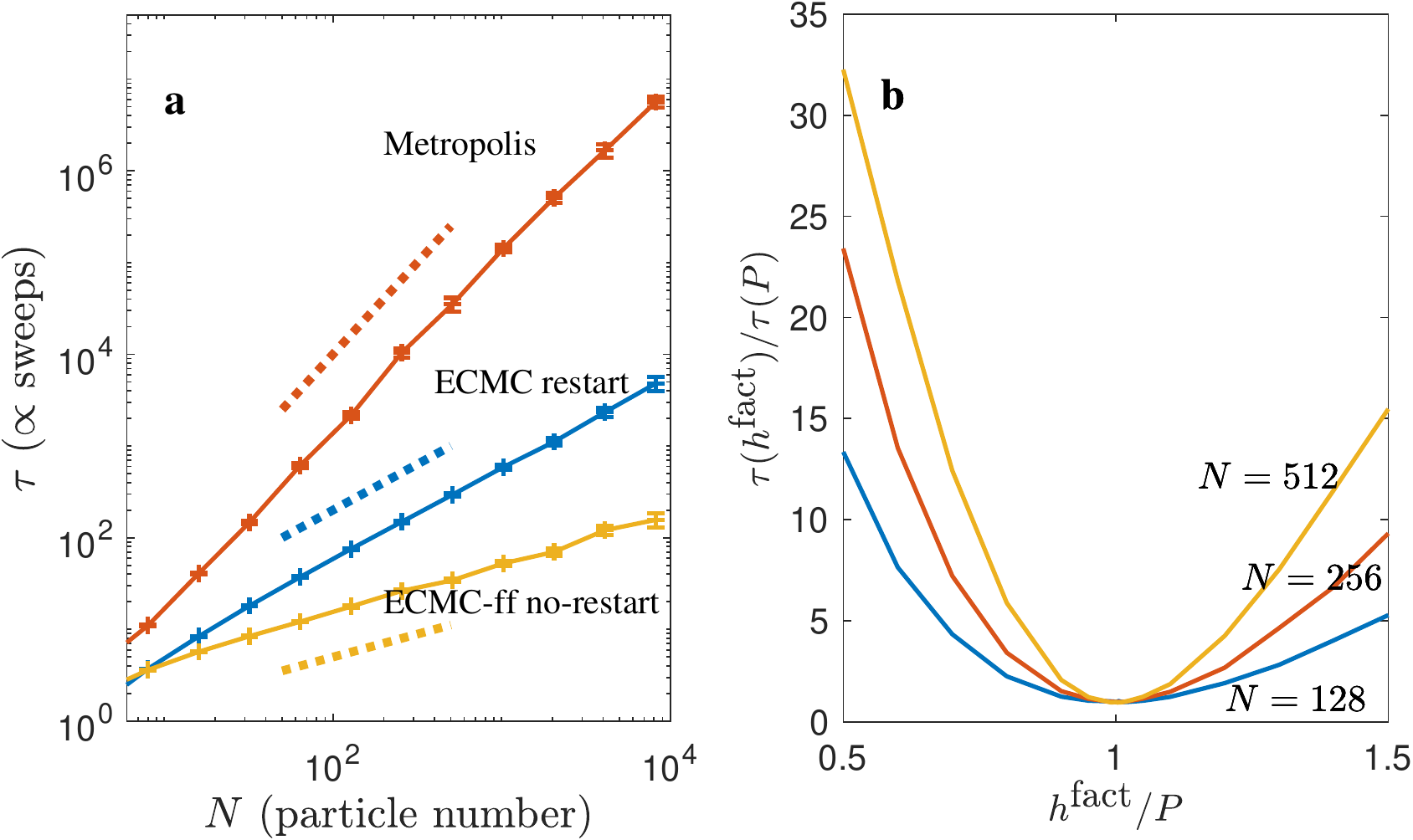}
\caption{Autocorrelation time $\tau$ (in sweeps) for 1D hard spheres. \subcap{a}
Reversible local Metropolis MCMC, ECMC (with restarts) and without
factor field and ECMC (without restarts) with optimal factor field,
$\FFeq$. Scalings $\tau \sim N^{1/2}$, $\tau \sim N$, and $\tau \sim N^2$
are indicated with dotted lines. \subcap{b} $\tau$ from ECMC \emph{vs.} factor
field  (ECMC, without restarts).
}\label{fig:Tonks_synopsis} 
\end{figure}

We study the autocorrelation time in sweeps (see \fig{fig:Tonks_synopsis}{a})
and compare with the reversible local Metropolis algorithm as well as ECMC
without a factor field. Again, we note the acceleration brought by the 
addition 
of a factor field with a dynamic exponent $z=1/2$, just as for the linear 
Lennard-Jones and the harmonic models. Non-optimal
factor fields slow down the dynamics of the longest wavelength 
modes, an 
effect which becomes stronger for larger $N$ (see \fig{fig:Tonks_synopsis}{b}).

\subsection{Evaluation of mixing times}

We have so far considered the equilibrium autocorrelation time, which is only
one of the two relevant measures for the speed of an algorithm; it measures the
time to move from one configuration (taken in equilibrium) to another 
independent one. The mixing time, in contrast, considers the time it takes to 
reach a first equilibrium
configuration from an arbitrary non-equilibrium state. The scaling with $N$ of
the equilibrium autocorrelation time and of the mixing time differs for many
MCMC algorithms in 1D particle systems (see~\cite{Levin2008} for a mathematical
discussion of mixing times and equilibrium autocorrelation times,
and~\cite{Lei2018b} for a discussion in the context of ECMC.)

\begin{figure}[htbp]
  \includegraphics[width=\linewidth]{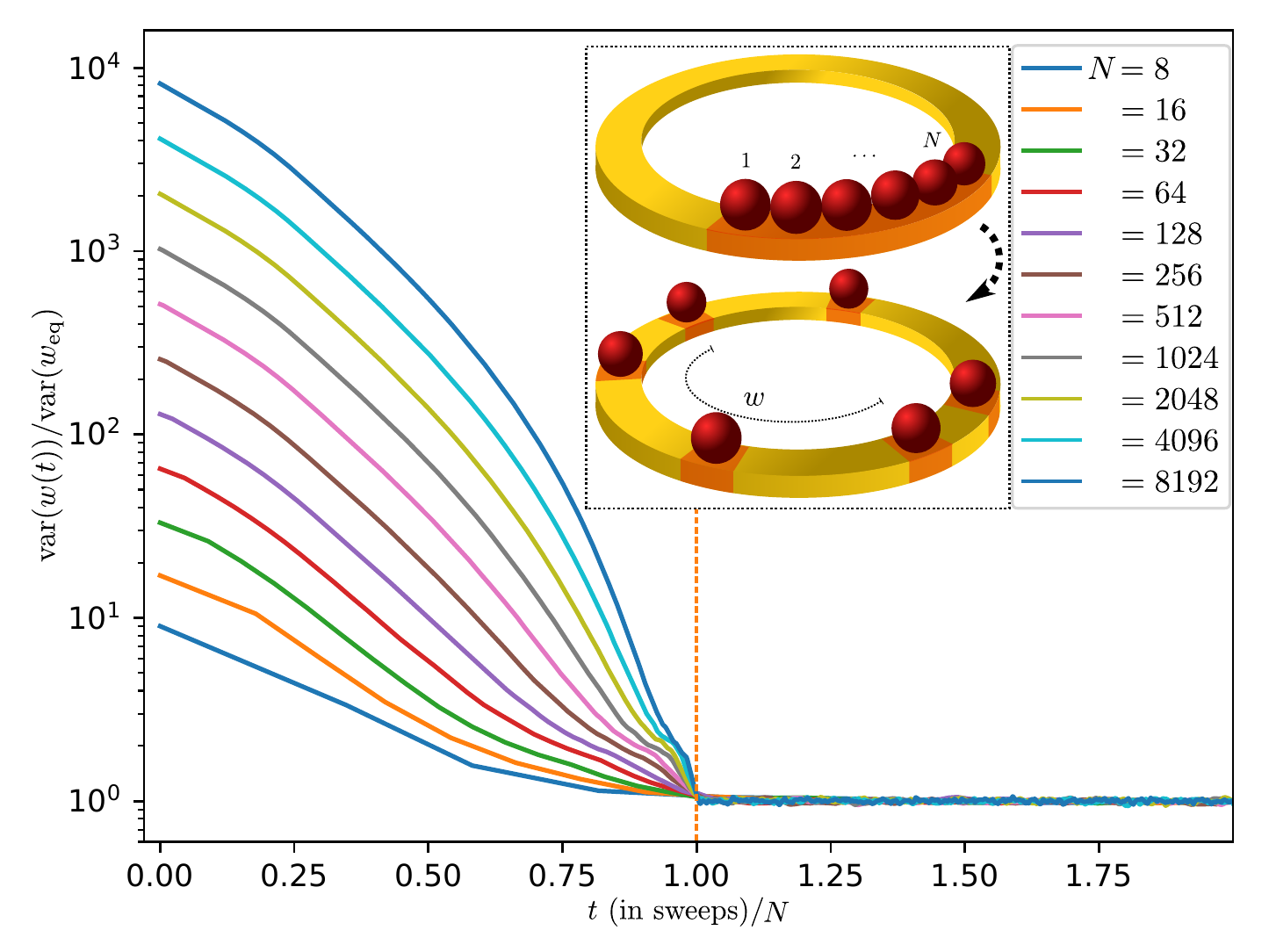}
  \caption{Variance of the half-system distance $w$
   (see \eq{equ:half}) \emph{vs.} time for various $N$ (hard-sphere model with 
   factor fields (without restarts)). The observable relaxes to its equilibrium 
   value at the mixing time $ (1.000 \pm 0.005) \times N$ sweeps for the 
    hard-sphere model with factor fields (without restarts). 
  }\label{fig:HarddiskMixing}
\end{figure}

To determine the mixing time for the hard-sphere model with factor fields, we
use a discretized version of the smallest Fourier coefficient of the structure
factor in \eq{equ:structure-factor}, namely the variance $\text{var}(w)$  of 
the \quot{half-system distance}
\begin{equation}
w =  x_{i+N/2} - x_i - N \sigma /2
  \label{equ:half}
\end{equation}
from a compact initial configuration~\cite{KapferKrauth2017} where 
$\text{var}(w) \propto N^2$ to the (exactly known) equilibrium value, which is 
$\propto N$.
Tracking the variance signals a mixing time very close to $N$ sweeps,
a value that we conjecture to be exact (see \fig{fig:HarddiskMixing}).  This
is a faster scaling than the $\bigOb{N\log N}$ sweep mixing-time behavior of 
ECMC
with restarts (without factor field)~\cite{Lei2018b}.

Relaxation occurs in the following manner from a compact configuration: First,
the active particle  is driven to the right end of the system which 
over-relaxes (see \fig{fig:HarddiskMixing}). This
drives the activity back into the bulk, to the boundary with the compact
interior. A series of cycles of increasing amplitude relaxes the
end of the system with penetration into the compact region following a law in
$\sqrt{t}$.  We note that the mixing time is longer than the 
equilibrium autocorrelation time (see the discussion in \sect{sec:DynAct}).

\begin{figure}[htbp]
  \includegraphics[width=0.9 \linewidth]{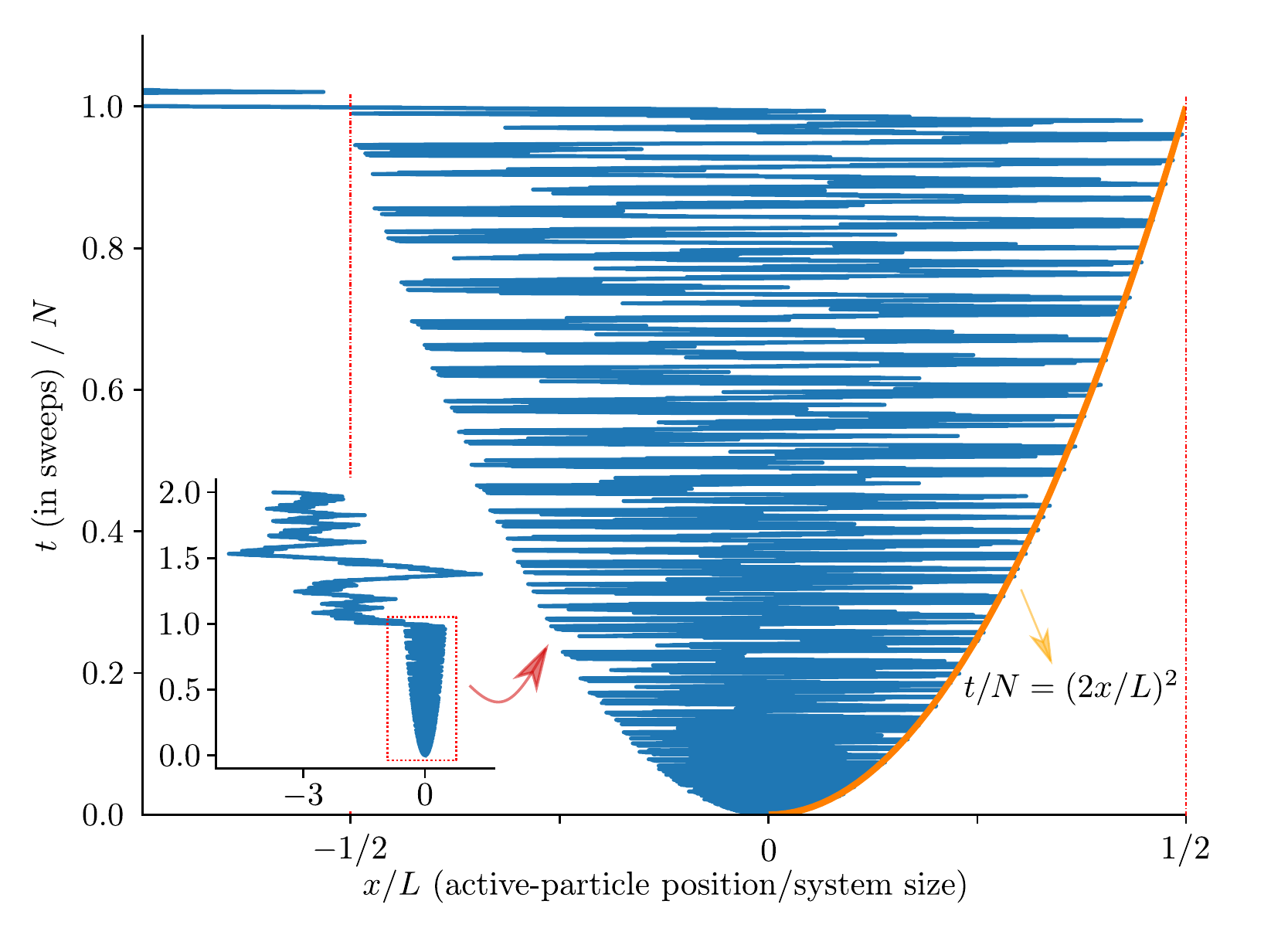}
  \caption{Position $x_i$ of the active sphere $i$ \emph{vs.} time $t$
    ($N = 4096, \sigma = L/(2N)$). At $t = 0$, the interval $[-L/2,0]$ is close
    packed, the interval $[0,L/2]$ is empty. The physical extent expands
    through oscillations, growing as $\sqrt{t}$, and reaches
    $[x_1, x_N] \simeq L$ at $t \simeq N$.  The inset illustrates the position
    of $x_i$ (without periodic wrapping) on a larger time interval.  }
\end{figure}
   
\section{Active-particle dynamics}\label{sec:DynAct}

\begin{figure}[htbp]
\includegraphics[width=\linewidth]{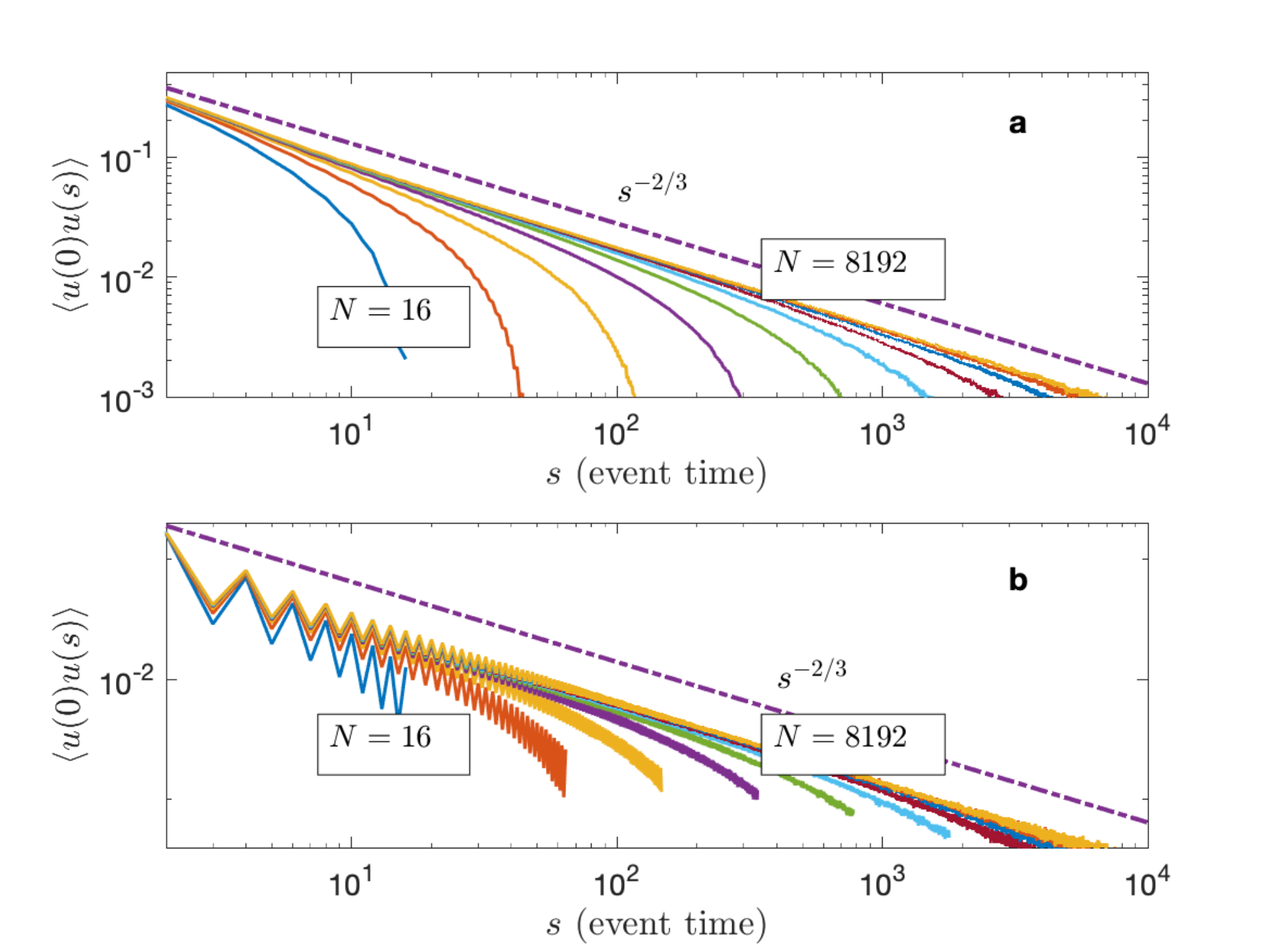}
\caption{Equilibrium autocorrelation of event steps $u \in \SET{-1,1}$ with 
event time
$s$ for ECMC with factor field (no restarts). \subcap{a} 1D Lennard-Jones
model shows monotonic decay. \subcap{b} 1D hard spheres display oscillatory
behavior with a power-law envelope. The scaling $ \mean{u(0)u(s)} \sim
s^{-2/3}$ is indicated with dotted lines (see \eq{equ:gamma}).
}\label{fig:P5R}
\end{figure}

The choice of factor fields, even if it is without incidence on 
spatial correlation functions and thermodynamic
properties at constant $L$, strongly influences the ECMC dynamics. In this 
section, we consider the large-scale motion of the particle $i(t)$ that is 
active at time $t$, in order to probe how the exponent $z=1/2$ arises from the 
local active-particle dynamics. It is convenient to take into consideration 
discrete \quot{event times} $s = 0,1,2,\dots$, rather than the continuous time 
$t$ of the Markov process. Because of the ordering of indices, we have
$i(s+1) = i \pm 1$, with \quot{event steps} $u(s)$ as $i(s+1) = i(s) + u(s)$ 
and $u(s) \in \SET{-1,1}$. It follows from \eq{equ:ecmcPressure} that $\mean{u} 
> 0$ and $\mean{u} < 0$ for  $P>0$ and $P< 0 $, respectively, which means that 
the ECMC trajectory is described by a forward drift (for $P>0$) or a backward 
drift (for $P<0$). With a factor field equal to $P$,  the drift terms 
vanish, and  ECMC trajectories feature positive and negative event steps 
(liftings $i(s+1)=i(s)+1$ and $i(s+1)=i(s)-1$) with equal 
probabilities~\cite{Anomalous2017}.  To better characterize the time series 
$u(s)$ in this case, for both hard spheres and Lennard-Jones particles, we 
compute the event-step autocorrelation $\mean{u(0) u(s)}$
(see \fig{fig:P5R}). We find that for large $N$, the autocorrelation decays as a 
power law:
\begin{equation}
  \mean{u(0) u(s)} \sim s^{-\gamma}.
\label{equ:gamma}
\end{equation}
(This scaling applies on times shorter than those required to explore the whole 
system. On longer time scales the correlation in \eq{equ:gamma} decays 
exponentially.)

The active particle at event time $s$ (without periodic wrapping) is given by
\begin{equation}
   i(s)=i(s=0) + \sum_{s'=1}^s u(s').
 \end{equation}
We now follow a trajectory which starts with $i(s=0)=0$.  For
vanishing long-range correlations in the event steps $u(s)$, the motion of the
activity, characterized by the second moment of $i(s)$, would be diffusive
($\mean{i^2(s)} \sim s$. Rather, we find for large $s$, using
\eq{equ:gamma} with $\gamma<1$:
\begin{equation}
 \mean{i^2(s)}  =\sum_{s'=1}^s \sum_ {s''=1}^s\mean{u(s') u(s'')} \sim  
s^{2-\gamma}.
\end{equation}
The position of the active particle is thus characterized by  super-diffusive
behavior. The observed value  $\gamma=2/3$ (see \fig{fig:P5R}) implies
\begin{equation}
\mean{i^2}  \sim s^{4/3}\quad \text{or}\quad |i| \sim s^{2/3}.
\end{equation}
The dynamics of the active particle has long-time memory for $N \to \infty$.
The trajectories contain long runs separated by changes of the direction of
motion, so that the average motion is undirected, as required by
\eq{equ:ecmcPressure}.

\subsection{Scaling for the active-particle 
dynamics}\label{sect:ScalingActiveParticle}

The discrete event time $s = 0,1,2,\dots$ grows with the time $t$ of the Markov
process (that we measure in sweeps) as
$s \propto N t$.  The same argument applies to the autocorrelation event time,
in events, and the autocorrelation time $\tau$, in sweeps,
$s_{\text{auto}} \propto N \tau(N)$.  The super-diffusive motion constrains the
dynamic exponent $z$ which relates complexity to system size:
\begin{equation}
  s_{\text{auto}} \sim N^{(1+z)}.
\end{equation}
A configuration can decorrelate from its previous history only if the
super-diffusive walk visits each sphere at least once. Thus we require:
\begin{equation}
  |i(\tau)| \sim {s_{\text{auto}}}^{1-\gamma/2} \sim 
N^{(1+z)(1-\gamma/2)} \ge N,
\end{equation}
implying that $z \ge \gamma/(2-\gamma) $. If we take $\gamma = 2/3$, we find
$z \ge 1/2$ compatible with the autocorrelation scaling reported previously
for the harmonic model~\cite{Lei2018b}, and also compatible with the data in
\figtwo{fig:P1L}{fig:Tonks_synopsis}. 

A supplementary physical hypothesis of perfect local equilibration during the
ECMC motion leads to a definite prediction for $\gamma$: the
equalibrium fluctuations in particle separation in a system section of length 
$|i|$
increase as
\begin{equation}
  \Delta x_p \sim |i|^{1/2}.
\end{equation} 
After $s$ events the active label visits particles in a volume $|i| 
\sim s^{1-\gamma/2}$, so that on average each particle moves
\begin{equation}
  \Delta x_\gamma =   \frac{s}{|i|} \sim s^{\gamma/2} \sim
  |i|^{\gamma/(2-\gamma)}
\label{equ:shift}
\end{equation} 
times. If we assume that the motion of the particles is comparable to that
required to resample the internal states of the section of length $|i|$ we find
$\Delta x_p \sim \Delta x_\gamma$ so that $\gamma = 2/3$, and $z=1/2$.

For this mechanism %(which leads to both $\gamma = 2/3$ and $z=1/2$)
to work, 
the 
correlated random motion of the active particle must
behave in a special way: both the mean and the standard deviation
of the distribution of $\Delta x_{\gamma}$ must have identical scaling with $s$.
(If only the mean increases the spheres will be displaced uniformly without
re-equilibrating the internal degrees of freedom.)

\subsection{Active-particle return probabilities}

The  distribution of \eq{equ:shift}) 
allows for a rapid decay of autocorrelation 
functions. We consider the dynamics of a particle which is active at time 
$s=0$. 
This particle can only move forward a large distance if the active label returns 
to it frequently, that is, if for many times $s$, one has $i(s)=i(s=0)$. 
We thus study in greater detail the returns to the origin of the active label, 
in the presence of factor fields.

\begin{figure}[htbp]
\includegraphics[width=0.9 \linewidth]{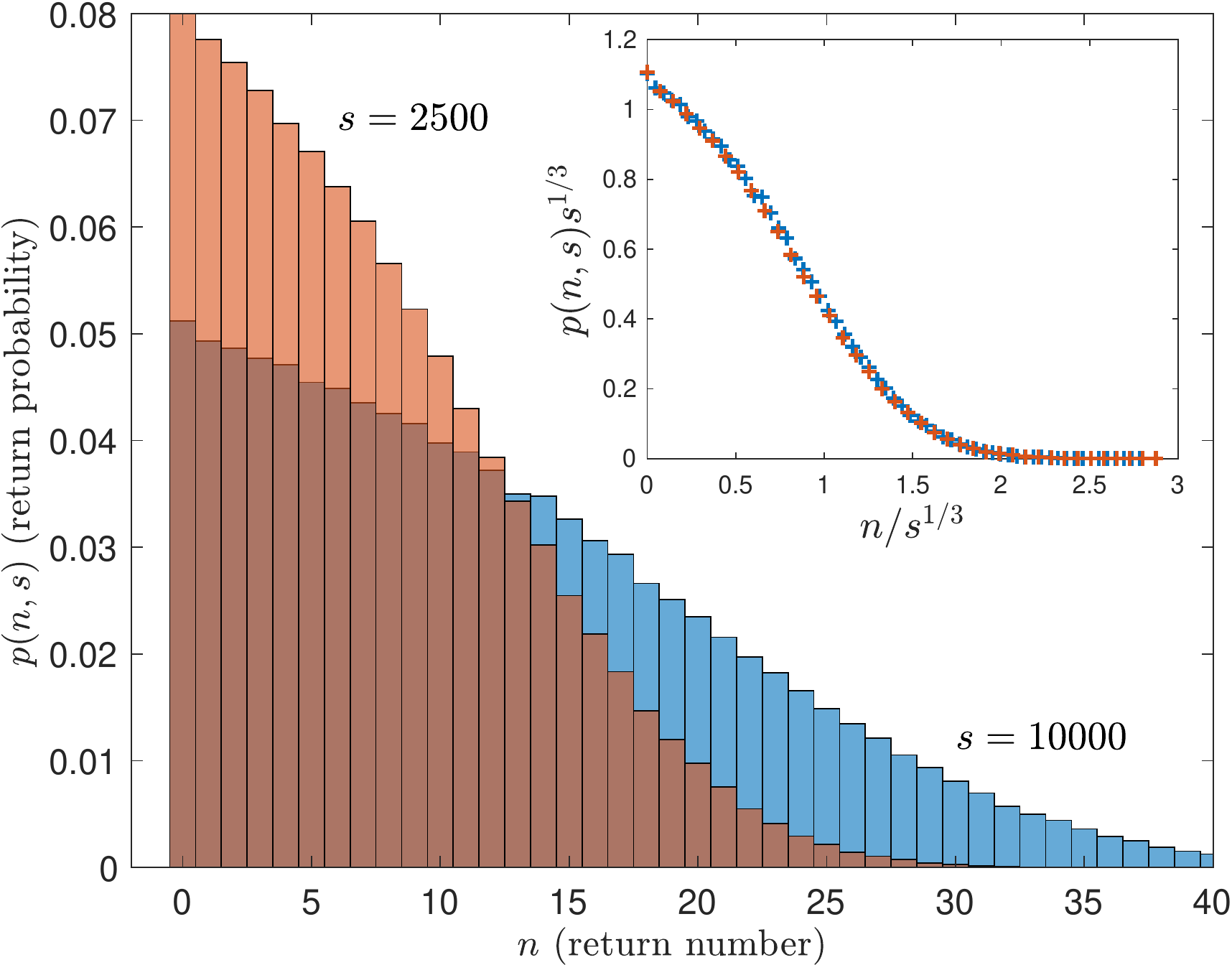}
\caption{ECMC with  factor fields for a 1D Lennard-Jones system ($N=8192$,
$T/ \epsilon=1$).  Probability $p(n,s)$ to return $n$ times to the original
active particle during $s$ events (for $s=2500$ and $s=10000$). Mean and
standard deviation of $p(n,s)$ both grow as $s^{\gamma/2}\sim s^{1/3}$ (see
\eq{equ:shift}). Inset: data collapse using scaling variables $p(n,s) s^{1/3}$
\emph{vs.} $n/s^{1/3}$.
}\label{fig:return}
\end{figure}

We generate an equilibrated configuration of the
Lennard-Jones system and from the signal $i(s)$ calculate the distribution of
the number $n$ of returns to the origin within $s$ events (see 
\fig{fig:return}).
For Brownian walks of length $s$, $n$ is related to the \quot{local
time}~\cite{feller-vol-1,localtime}, and the local-time distribution $p(n,s)$
is half-gaussian defined for $n > 0$.  In ECMC, the probability $p(n,s)$ of returns of
the active-particle label to the origin (which gives the number of forward
steps)  is also maximum at zero, and decays monotonically
with $n$. The mean and standard deviation of the number of steps drawn from such
a distribution grow in the same way with $s$ (see inset of \fig{fig:return}).
Even though the whole system moves forward in an ECMC simulation the dynamics is
spatially heterogeneous. Widely separated particles move forward with different
numbers of steps so that the internal modes of the system are efficiently
resampled as is needed for the ansatz in \eq{equ:shift} to apply.
  
\section{Conclusions}

We have compared in detail the dynamics of three simulation
methods (reversible MCMC, MD and ECMC) for 1D systems with local interactions. 
We have shown that in many
situations ECMC displays the same dynamic scaling ($z=1$) as molecular
dynamics. Both are asymptotically faster than the diffusive behavior
found in MCMC ($z=2$). With a good choice of factors, ECMC is  much faster than
MD, since it does not need to use a small integration time step to stably
explore configurations. Furthermore, unlike MD, ECMC is exact to 
machine precision, as it is free from time-discretization errors. 

Generalizing from the 1D harmonic model, we map 1D systems
onto thermodynamically equivalent systems at zero pressure with
periodic boundary conditions.
This leads to further acceleration of ECMC for both smooth and discontinuous
potentials. We have found in this case a remarkably low dynamic 
exponent
($z=1/2$), better than MD. This acceleration is associated
with a modification of the dynamics of the event steps (as a consequence
of \eq{equ:ecmcPressure}). Rather than displaying directed motion, the
signal $i(s)$ becomes super-diffusive and optimally explores local density
fluctuations, being driven forward in regions of high density, and back in
regions of low density. A scaling hypothesis predicts a
super-diffusive law of the form $\langle i^2(s)\rangle \ \sim s^{4/3}$ for the
dynamics of the active label as well as an explanation for the emergence of the
exponent $z=1/2$.

There is a clear interest in generalizing these results to higher-dimensional
models. Already a two and three-dimensional harmonic model has been
shown~\cite{Lei2018b} to display accelerated convergence in the ECMC
algorithm. In geometries which remain fixed, such as the $XY$ model or fixed
harmonic networks (without disorder) it appears possible to implement
generalized factor fields.  With fluctuating neighbor relations, for 
instance in a fluid, the generalization of factor fields will represent an 
interesting challenge
\cite{Wallace70}.

\acknowledgments{}
W.K. acknowledges support from the Alexander von Humboldt Foundation.

\bibliographystyle{ieeetr}
\bibliography{General,../Factsheet/LennardJones}
\end{document}